\documentstyle[twocolumn,prl,aps]{revtex}
\begin{document}
\draft
\title{D-wave Bose-Einstein condensation and the London penetration depth
in superconducting cuprates.}
\author{A.S. Alexandrov \and R.T. Giles}
\address{Department of Physics, Loughborough University, Loughborough, LE11
3TU, U.K.}
\date{\today}
\maketitle

\begin{abstract}
We show that bipolaron formation
leads to a d-wave  Bose-Einstein condensate in cuprates.
It is the bipolaron energy dispersion  rather than a particular
pairing interaction which
is responsible for the  $d$-wave symmetry. The unusual low-temperature
dependence of the magnetic field penetration depth~$\lambda(T)$ in cuprates is
explained by the localisation of bosons in the
random potential.  Both linear positive and negative slopes of
 $\lambda(T)$ are occur depending
on the random field profile.

\end{abstract}

\pacs{74.20.-z,74.20.Mn}

The evidence for a $d$-like order parameter (changing sign when the
$CuO_{2}$ plane is rotated by $\pi/2$) has been reviewed by Annett,
Goldenfeld and Legget \cite{ann} and more recently by Brandow \cite{bra}
and by several other authors. A number of phase-sensitive
experiments \cite{pha}
provide unambiguous evidence in this direction; furthermore, the
low temperature magnetic  penetration depth \cite{bon,xia} has been found to be
linear in many cuprates as expected for
a d-wave BCS superconductor.  However, $SIN$
tunnelling studies \cite{deu,ren}
and some high-precision magnetic measurements \cite{gol}
show the more usual $s$-like shape of
the gap function
or even  reveal an upturn in the temperature dependence of the
penetration depth below some characteristic temperature \cite {wal}.
   One can reach a compromise between conflicting experimental
 results  by mixing
 $s$ and $d$ order parameters  or invoking
 `anomalous Meissner currents' due to
 surface-induced bound states (often violating time reversal
 symmetry).  However, the observation of the normal state pseudogap in
 tunnelling (STM) and photoemission spectra (ARPES),  non Fermi-liquid
 normal state kinetics and thermodynamics, and  unusual critical
 phenomena tell us that many high-$T_{c}$
 cuprates are not $BCS$ superconductors  \cite{alemot}. In particular, both
ARPES
 \cite{she} and STM \cite{ren} experiments have shown that the maximum
 energy gap is more than three times larger than that expected from the
 d-wave
 BCS theory and persists into the normal state irrespective of doping.
 The gap as well as other major features of STM and ARPES have been
 recently explained with bipolarons \cite{ale2}.
Comparison of tunnelling and Andreev gap determinations on
 yittrium, lanthanum and bismuth-based cuprates \cite{deu} at various
 doping levels have unambigously revealed the existence of two energy scales
as  expected for the bipolaronic superconductors\cite{deu2}.
Hence an explanation for the
 $d$-like order parameter and the anomalous penetration depth
  should be found independent of the $BCS$ gap equation.

While in a BCS superconductor all energy scales and symmetries are
strictly identical,
the symmetry of the Bose-Einstein condensate in  the bipolaronic superconductor
 should be distinguished from that of the
`internal' wave function of a single bipolaron.
 In this letter we show
 that the Bose-Einstein condensate in cuprates is $d$-wave
owing to the
bipolaron energy band structure  rather than to a particular
pairing interaction (see also \cite{ale2}), while
the low temperature dependence of the
penetration depth, $\lambda(T)$ is
determined by the localisation of bipolarons.
 Both linear positive and negative slopes of
 $\lambda(T)$ occur depending
on the random field profile.

Consideration of perovskite lattice structures shows that small
inter-site bipolarons are perfectly mobile even if the electron-phonon
 coupling is strong and the bipolaron binding energy is large\cite{chak}.
Different
 bipolaron configurations can be found  with computer
simulation techniques based on the minimization of the ground state
energy of an ionic lattice with two holes. The intersite pairing of
the in-plane oxygen hole  with the  $apex$ one is
energetically favorable in the layered perovskite structures as
 established by
Catlow $et$ $al$ \cite{cat}.
This apex or peroxy-like bipolaron can tunnel from one cell to another  via a
direct $single$ $polaron$ tunnelling from one apex oxygen to its apex
neighbor.
 The bipolaron band structure has been derived by one of us \cite{ale}
as
\begin{equation}
E^{x,y}_{\bf k}=tcos( k_{x,y})-t'cos( k_{y,x}).
\end{equation}
Here the in-plane lattice constant is taken as $a=1$,
 $t$ is twice the bipolaron
 hopping integral between $p$ orbitals of the same symmetry
elongated in the direction
 of the hopping ($pp\sigma$)
 and $-t'$ is twice the hopping integral in the perpendicular
 direction ($pp\pi$).  The  bipolaron energy spectrum in  the tight
binding approximation consists of two bands $E^{x,y}$ formed by the
overlap of
$p_{x}$  and  $p_{y}$ $apex$ $polaron$ orbitals, respectively.
The energy band minima are found at the Brillouin zone
 boundary,  ${\bf
 k}=(\pm \pi,0)$ and  ${\bf k}=(0, \pm \pi)$ rather than at the  $\Gamma$ point
 owing to the opposite sign of the $pp\sigma$ and $pp\pi$
 hopping integrals.  Only their relative sign  is important, and we choose
$t,t' >0$.
Neither band is 
invariant under crystal symmetry but the degenerate doublet is
an irreducible representation; under a $\pi/2$ rotation the $x$ band
transforms into $y$ and vice versa.

If the bipolaron  density  is low, the
 bipolaron Hamiltonian can be mapped onto the charged Bose-gas
 \cite{alemot}. Charged bosons are condensed below $T_{c}$ into the
 states of the Brillouin zone with the lowest energy, which are ${\bf
 k}=(\pm \pi,0)$ and  ${\bf k}=(0, \pm \pi)$ for the $x$ and $y$
 bipolarons, respectively. These four states are degenerate, so the
 order parameter $\Psi({\bf m})$ (the condensate wave function) in the real
(site) space ${\bf m}= (m_{x},m_{y})$  is given
 by
 \begin{eqnarray}
 \Psi_{\pm}({\bf m}) &=& N^{-1/2}\sum_{{\bf k}=(\pm \pi,0),(0,\pm \pi)} b_{\bf
k} exp(-i{\bf k \cdot m})\nonumber\\
 &=& n_c^{1/2} \left[ \cos (\pi m_{x}) \pm \cos
 (\pi m_{y})\right] \label{eq:psi},
 \end{eqnarray}
 where $N$ is the number of cells in the crystal, $b_{\bf k}$ is the
 bipolaron
(boson) annihilation operator in
 ${\bf
 k}$ space (which is a $c$-number for the condensate), and $n_c$ is the
 number of bosons per cell in the condensate.
 Other combinations of the four degenerate states do not respect
 time-reversal and (or) parity symmetry.
The two solutions, Eq.(\ref{eq:psi}),  are physically identical being
 related by:
 $\Psi_{+}(m_{x},m_{y})=\Psi_{-}(m_{x},m_{y}+1)$. They have
 $d$-wave  symmetry changing  sign when the
$CuO_{2}$ plane is rotated by $\pi/2$ around $(0,0)$  for $\Psi_{-}$ or
around (0,1) for $\Psi_{+}$ (Fig.1).
  The
$d$-wave symmetry is entirely due to the bipolaron energy dispersion
with four minima at ${\bf k \neq 0}$. With
the energy minimum located at the $\Gamma$ point  of the Brillouin zone the
condensate is $s$-like.

If the total number of bipolarons in one unit cell
is $n$ of which $n_L$ are in localised states and $n_D$ are in
delocalised states then the number in the condensate~$n_c$ is
\begin{equation}
  n_c=n-n_L-n_D
\end{equation}
and the London penetration depth~$\lambda\propto 1/\sqrt{n_c}$.
Taking the delocalised bipolarons to be a free three-dimensional gas
we have $n_D\propto T^\frac{3}{2}$.  Here we use a simple model we
have presented previously\cite{alegil} to calculate the temperature
dependence of $n_L$ and find that at low temperature $n_L$ varies
linearly with temperature.  Thus in the limit of low temperature we
can neglect $n_D$ and make the approximation
\begin{equation}
  n_c\approx n-n_L
\end{equation}
In this limit $\lambda(T)-\lambda(0)$ is small and so
\begin{equation}
  \lambda(T)-\lambda(0) \propto n_L(T)-n_L(0) \label{eq:lambda}
\end{equation}
i.e. $\lambda$ has the same temperature dependence as $n_L$.

The  picture of interacting bosons
filling up all the localised single-particle states in a random potential
and Bose-condensing into the first extended state is  known in the
literature \cite{her,ma,fis,fis2}.
The comprehensive scaling analysis of neutral \cite{fis}
and  charged
bosons  \cite{fis2} allows us to describe
the quantum Bose glass-superfluid transition while the thermodynamics of each
phase away from the
transition can be studied with the physically plausible models of neutral
\cite{gun}
and charged bosons \cite{alegil} in a random potential. These models
are based on a separation of
localised single-particle states from delocalised states.
The  renormalisation of the
 single-particle energies  by the collective mode does not
affect the temperature dependence of any of the thermodynamic
functions at low temperatures \cite{alegil}.
Hence,  one assumes that at some temperature~$T_c$ bosons are condensed at the
mobility edge~$E_c$ so that the chemical potential $\mu=E_c$, and
some of the bosons are in localised states below the mobility edge.

For
convenience we choose $E_c=0$. When two or more charged bosons
are in a single localised state of energy $E$ there
may be significant Coulomb energy and we  take this into account as follows.
 The localisation length
$\xi$ is thought to depend on $E$ via
\begin{equation}
  \xi\propto \frac{1}{\left(-E\right)^\nu}
  \label{eq:xi}
  \end{equation}
  where $\nu>0$. The Coulomb potential energy of $p$ charged bosons
confined within a
radius $\xi$ can be expected to be
\begin{equation}
   \mbox{potential energy }\sim\frac{p(p-1)e^2}{\epsilon_0\xi}.
   \label{eq:pe}
\end{equation}
where $\epsilon_0$ is the dielectric constant.
Thus the total energy of $p$ bosons in a localised state of energy
$E$ is taken to be
\begin{equation}
  w(E)=pE + p(p-1)\kappa\left(-E\right)^\nu
  \label{eq:w}
\end{equation}
where $\kappa>0$.  We can thus define an energy scale $-E_1$:
\begin{equation}
  -E_1 = \kappa^{\frac{1}{1-\nu}}.
\end{equation}
From here on we choose our units of energy such that $E_1=-1$.
We take the total energy of  charged bosons in localised  states to be
the sum of the energies of the bosons in the individual potential
wells.  The partition function $Z$ for such a system is then the
product of the partition functions $z(E)$ for each of the wells,
\begin{eqnarray}
   z(E)  =  e^{\alpha p_0^2}
       \sum_{p=0}^{\infty} e^{-\alpha (p-p_0)^2}
                \label{e:zls}
\end{eqnarray}
where
\begin{equation}
  p_0=\frac{1}{2}\left\{1+(-E)^{1-\nu}\right\}
\end{equation}
\begin{equation}
  \alpha=\frac{(-E)^\nu}{\theta}.
\end{equation}
and
\begin{equation}
  \theta=\frac{k_BT}{(-E_1)}
\end{equation}

The average number $n_L$ of bosons in localised states is
\begin{equation}
  n_L = \int_{-\infty}^{0} dE \left<p\right> \rho_L(E) \label{eq:N_L}
\end{equation}
where the mean occupancy $\left<p\right>$ of a single localised state
is taken to be
\begin{equation}
\left<p\right>  =  \frac{\sum_{p=0}^{\infty}p\,
  e^{-\alpha(p-p_0)^2}} {\sum_{p=0}^{\infty}e^{-\alpha(p-p_0)^2}}
  \label{eq:p}
\end{equation}
and $\rho_{L}(E)$ is the one-particle density of localised states
per unit cell below the mobility edge, which is taken to be
\begin{equation}
  \rho_{L}(E) = \frac{N_L}{\gamma}e^{\frac{E}{\gamma}}.
  \label{eq:rho}
\end{equation}
We now focus on the temperature dependence of $n_L$ at low temperature
($\theta\ll 1$) for the case where the width of the impurity tail
$\gamma$
is large ($\gamma>1$).
 In
the following  we consider first the case $\nu>1$ and then $\nu<1$.

If $\nu>1$ we can approximate $n_{L}$ as
\begin{equation}
   \frac{n_L}{N_L}  \approx  1 +
   \frac{\nu-1}{2(2-\nu)\gamma}
  +
   \frac{2\theta}{(2-\nu)\gamma}\ln 2 \label{eq:nu>1}
\end{equation}
So we expect $n_L$ to be close to the total number of wells $N_L$ and
to increase linearly with temperature.
Fig~2a compares this analytical formula
with accurate  numerical calculation for the case $\nu=1.5$, $\gamma=20$.
We also note that even when $\gamma<1$, $n_L(\theta)$ will still be
linear with the same slope provided that $\theta\ll\gamma$.

If $\nu<1$ we obtain, keeping only the lowest power of $\theta$ (valid provided
$\theta^\frac{1}{\nu}\ll\theta$)
\begin{equation}
 \frac{n_L}{N_L} = \frac{1}{2}+
\frac{\Gamma(2-\nu)\gamma^{1-\nu}}{2}+\frac{1-\nu}{2(2-\nu)\gamma}
-\frac{\theta}{\gamma}\ln
2 \label{eq:nu<1}
\end{equation}
Hence in this case $n_L$ decreases linearly with increasing temperature (in the
low temperature limit).
Fig~2b compares this analytical formula
with the numerical calculation for the case $\nu=0.65$,
$\gamma=20$.

Fig~3 shows that the low temperature experimental data \cite{wal} on the
London penetration depth~$\lambda$ of $YBCO$ films can be fitted very well by
this theory with $\nu<1$.  It is more usual to see $\lambda$ increase
linearly with temperature \cite{bon,xia} and this would correspond to $\nu>1$.

The exponent $\nu$ depends on the random field profile. We believe
that $\nu <1$ is more probable for a rapidly varying random
potential while  $\nu >1$ is more likely for a slowly varying one. Both $\nu <1$ and
 $\nu >1$ are observed in doped semiconductors.
 Hence, it is not surprising that drastically different low-temperature
dependence of the London
penetration depth is observed  in different samples of doped cuprates.
 In the framework of our approach $\lambda (T)$ is related to the
localisation of carriers at low temperatures rather than to any energy
scale characteristic of the condensate. The excitation spectrum of the
charged Bose-liquid determines, however,  the
temperature dependence of $\lambda(T)$ at higher
temperatures including an unusual critical behaviour
near $T_{c}$ \cite{alemot}.

 Many  thermodynamic, magnetic and kinetic properties of
cuprates have been understood in the framework of the bipolaron scenario
\cite{alemot}. We admit, however, that one experimental fact is
enough to destroy any theory. In particular, the single-particle
spectral function seen by ARPES \cite{she} was interpreted by several
authors as a
Fermi liquid feature of the normal state incompatible with bipolarons.
Most (but not all) of these measurements indicated a large Fermi
surface which one would think should evolve with doping as $(1-x)$ (where $x$ is
the number of holes introduced by doping) but such an evolution is in clear
contradiction with kinetic and thermodynamic measurements which show
an evolution proportional to $x$ . Only recently it has been
established that there is a normal state gap in ARPES and SIN
tunnelling, existing well above
$T_{c}$ irrespective of the doping level \cite{she,ren,bia}. The
`Fermi surface' shows missing segments just near the $M$ points\cite {bia}
where we
expect the Bose-Einstein condensation. A plausible explanation is that
there are two liquids in cuprates, the normal Fermi liquid and the
charged Bose-liquid (this mixture has been theoretically discussed a long time
ago \cite{alexme}). However, it is difficult to see how this scenario
could explain the doping dependence of  $dc$ and $ac $
conductivity or of the magnetic susceptibility and carrier specific
heat. On the other hand, the single-particle spectral function of a
pure bipolaronic system has been recently  derived by one of us \cite{ale2}.
It describes the  spectral features of tunnelling
and photoemission in cuprates, in particular, the temperature independent
gap and the
anomalous $gap/T_{c}$ ratio, injection/emission asymmetry both in
magnitude and  shape, zero-bias conductance at zero temperature, the
spectral shape inside and
outside the gap region, temperature/doping dependence and dip-hump
structure  of the tunnelling conductance and
ARPES. The zero-bias conductance and any
spectral weight at the chemical potential appear only  due to single
polaronic states  localised by the random  field. The model is thus
compatible with the doping evolution of thermodynamic and kinetic properties.

In conclusion we suggest, within
the framework of the bipolaron theory of cuprates~\cite{alemot},
explanations
 of the $d$ wave symmetry of the
ground state and the anomalous temperature dependence of the London
penetration depth,  compatible with the non Fermi-liquid normal state,
anomalous critical behaviour, and the ARPES and tunnelling data

We acknowledge illuminating discussions with J. Annett, A.R. Bishop, A.
Bussmann-Holder, E. Dagotto, G. Deutscher, V.V. Kabanov, D. Mihailovic,
K.A. M\"{u}ller, and C. Panagopoulos.

\begin{figure}
\caption{ D-wave condensate wave function,
$\Psi_{-}({\bf m})=n_c^{1/2} \left[ \cos (\pi m_{x}) -\cos(\pi
m_{y})\right]$ in the real (Wannier) space.
 The order parameter has
different signs in the shaded cells and is zero in the blank cells.}
\end{figure}
\begin{figure}
\caption{Dependence of the density of localised bosons $n_L$ on temperature
$\theta$ for $\gamma=20$.  (a) $\nu=1.5$, (b) $\nu=0.65$. The solid
lines correspond to the low temperature predictions from equations 
\ref{eq:nu>1} and \ref{eq:nu<1}, while the dashed lines are derived
from an accurate numerical
calculation.}
\end{figure}

\begin{figure}
\caption{Fit to the London penetration depth obtained by Walter 
$et$ $al$ [8] for a non-irradiated YBCO film. The parameter values
from the
fit were
$E_1=-74K$, $\gamma=20$ and $\nu=0.67$. }

\end{figure}

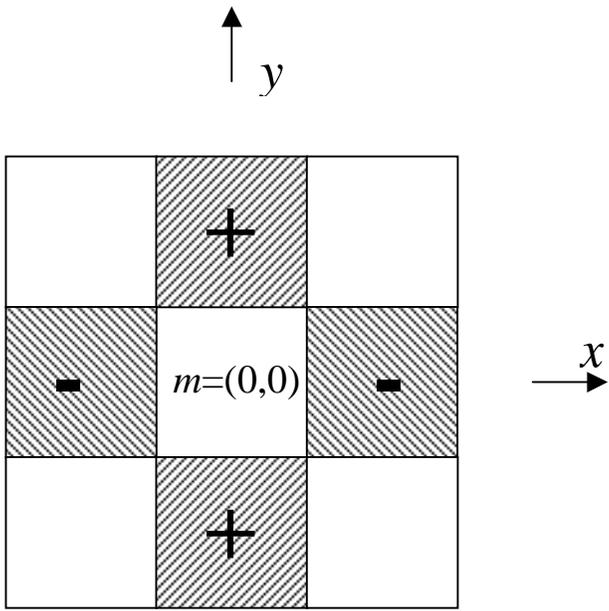
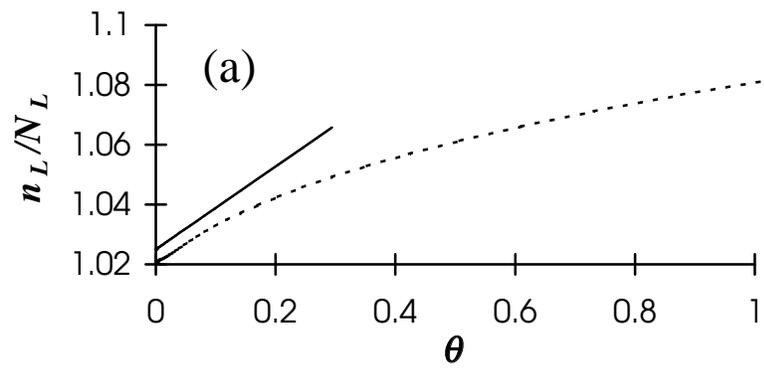
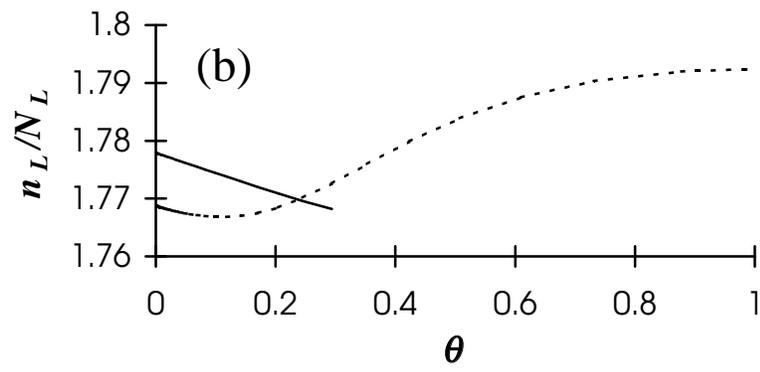
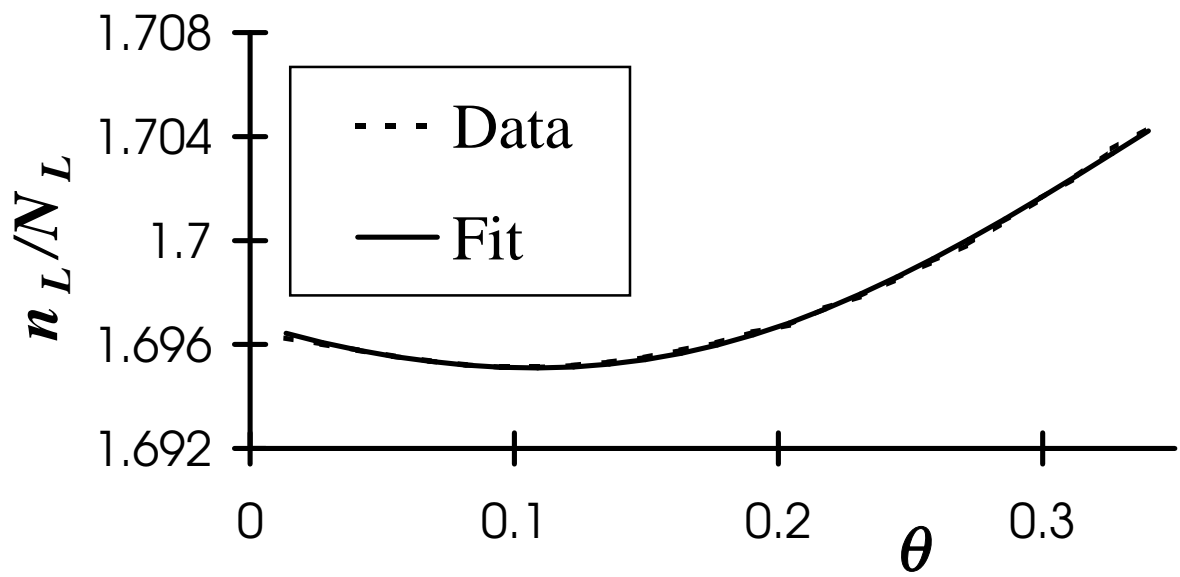
\end{document}